\documentclass[grl]{AGUTeX}

\usepackage{graphicx}
\usepackage{amsmath}
\usepackage{textcomp}
\usepackage{amssymb}
\usepackage{rotating}
\usepackage{lineno}
\usepackage[version=3]{mhchem} 
\def\D{\mathrm{d}}
\def\E{\mathrm{e}}

\usepackage{graphicx}

\setkeys{Gin}{draft=false}

\authorrunninghead{WORDSWORTH ET AL.}

\titlerunninghead{REDUCING GREENHOUSES ON EARLY MARS}

\authoraddr{Corresponding author: Robin Wordsworth, School of Engineering and Applied Sciences, Harvard University, Cambridge, MA 02138. (rwordsworth@seas.harvard.edu)}

\begin{document}


\title{Transient reducing greenhouse warming on early Mars}




\authors{R.~Wordsworth\altaffilmark{1,2},
Y.~Kalugina\altaffilmark{3}, 
S.~Lokshtanov\altaffilmark{4,5}, 
A.~Vigasin\altaffilmark{5}, 
B.~Ehlmann\altaffilmark{6,7}, 
J.~Head\altaffilmark{8}, 
C.~Sanders\altaffilmark{2,6} and
H.~Wang\altaffilmark{1}}

\altaffiltext{1}{School of Engineering and Applied Sciences, Harvard University, Cambridge, MA 02138, USA}
\altaffiltext{2}{Department of Earth and Planetary Sciences, Harvard University, Cambridge, MA 02138, USA}
\altaffiltext{3}{Department of Optics and Spectroscopy, Tomsk State University, Tomsk, Russia}
\altaffiltext{4}{Lomonosov Moscow State University, Chemistry Department, Moscow, Russia}
\altaffiltext{5}{Obukhov Institute of Atmospheric Physics, Russian Academy of Sciences, Moscow, Russia}
\altaffiltext{6}{Division of Geological and Planetary Sciences, California Institute of Technology, Pasadena, CA 91125, USA}
\altaffiltext{7}{Jet Propulsion Laboratory, California Institute of Technology, Pasadena, CA 91109, USA}
\altaffiltext{8}{Department of Earth, Environmental and Planetary Sciences, Brown University, Providence, RI 02912, USA}



\begin{abstract}
The evidence for abundant liquid water on early Mars despite the faint young Sun is a long-standing problem in planetary research. Here we present new \emph{ab initio} spectroscopic and line-by-line climate calculations of the warming potential of reduced atmospheres on early Mars. We show that the strength of both \ce{CO2}-\ce{H2} and \ce{CO2}-\ce{CH4} collision-induced absorption (CIA) has previously been significantly underestimated. Contrary to previous expectations, methane could have acted as a powerful greenhouse gas on early Mars due to \ce{CO2}-\ce{CH4} CIA in the critical 250-500~cm$^{-1}$ spectral window region. In atmospheres of 0.5~bar \ce{CO2} or more, percent levels of \ce{H2} or \ce{CH4} raise annual mean surface temperatures by tens of degrees, with temperatures reaching 273~K for pressures of 1.25-2~bar and 2-10\% of \ce{H2} and \ce{CH4}. Methane and hydrogen produced following aqueous alteration of Mars' crust could have combined with volcanically outgassed \ce{CO2} to form transient atmospheres of this composition 4.5-3.5~Ga. Our results also suggest that inhabited exoplanets could retain surface liquid water at significant distances from their host stars.
\end{abstract}



\begin{article}

\section{Introduction}

Today Mars is cold and dry, with annual mean surface temperatures of around $ -60^\circ$~C and a mainly arid, hyperoxidising surface. In the past, however, a diverse array of geological evidence points to episodically warmer and wetter conditions. This evidence includes dendritic valley networks distributed over large regions of the equatorial and southern Noachian highlands, fluvial conglomerates, open-basin lakes, and fluvolacustrine deposits \citep{Fassett2008a,Hynek2010,Grotzinger2015}.

This evidence for surface aqueous modification is paradoxical, because the Sun's luminosity was only around 75-80\% of its present-day value during the period 3-3.8~Ga when most of the erosion occurred. In combination with Mars' distant orbit, this implies cold surface conditions: even given a planetary albedo of zero, early Mars would have had an equilibrium temperature of only 210~K \citep{Wordsworth2016b}.  
Carbon dioxide provides some greenhouse warming but not enough: climate models that assume pure \ce{CO2}-\ce{H2O} atmospheres consistently predict global mean temperatures of less than 240~K for any surface pressure \citep{Kasting1991,Wordsworth2013a}. Many alternative mechanisms to warm early Mars have subsequently been investigated, including \ce{CO2} clouds \citep{Forget1997}, large meteorite impacts \citep{Segura2002}, sulfur dioxide emission from volcanos \citep{Postawko1986,Halevy2014}, and local snowmelt due to diurnal forcing and/or obliquity and eccentricity variations \citep[e.g.,][]{Wordsworth2013a}. However, all suffer shortcomings that render them unlikely as the main explanation \citep{Forget2013,Ramirez2014,Kerber2015,Wordsworth2016b}. 

Reducing greenhouse solutions for early Mars have also been considered previously. \citet{Sagan1977} argued that early Mars might have been warmed by a hydrogen-dominated atmosphere or by abundant \ce{NH3}. However, a hydrogen-dominated atmosphere would be lost to space rapidly after formation and \ce{NH3} is photolysed rapidly by UV radiation and lacks a plausible martian source. Later, in a paper focused on the early Earth, \citet{Wordsworth2013c} showed that hydrogen could act as an important greenhouse gas in terrestrial-type atmospheres even in abundances of a few percent, due to the strength of its collision-induced absorption in combination with heavier gases like nitrogen. \citet{Ramirez2014} applied this mechanism to early Mars, where they argued that \ce{H2} emitted from volcanoes into a \ce{CO2}-dominated atmosphere could have kept Mars in a `warm and wet' state for periods of 10s of millions of years or longer. However, lacking \ce{CO2}-\ce{H2} CIA data they used the same \ce{N2}-\ce{H2} data as \citet{Wordsworth2013c} for their climate calculations. As a result, they found that $>5$\% \ce{H2} in a 4~bar \ce{CO2} atmosphere (20\% \ce{H2} in a 1.3~bar atmosphere) was required to raise annual mean surface temperatures to the melting point of liquid water: an amount that is not consistent either with constraints on the total amount of \ce{CO2} present in the Noachian \citep{Hu2015} or estimates of the rate of hydrogen escape to space \citep{Ramirez2014}. Hence the early martian faint young Sun paradox remains unresolved.

Here we describe new spectroscopic and one-dimensional line-by-line climate calculations that we have performed to assess the warming potential of reducing climates on early Mars. We find \ce{CO2}-\ce{H2} warming to be significantly more effective than predicted by \citet{Ramirez2014} due to the strong polarizability and multipole moments of \ce{CO2}. Furthermore, we show for the first time that methane (\ce{CH4}) could have been an effective warming agent on early Mars, due to the peak of \ce{CO2}-\ce{CH4} CIA in a key spectral window region. {We propose that early Mars could have been transiently warmed by emission of these gases due to crustal aqueous alteration, volcanism and impact events. Our results also have implications for the habitability of exoplanets that orbit far from their host stars.}

\section{Methods}

To calculate the collision-induced absorption spectra for \ce{CO2}-\ce{CH4} and \ce{CO2}-\ce{H2} pairs, we first acquired the potential energy surface (PES) and induced dipole surface (IDS) for the relevant molecular complex. The PES for \ce{CO2}-\ce{H2} calculated at the coupled-cluster level was taken from the literature \citep{Li2010}. For the IDS for \ce{CO2}-\ce{H2} and both the PES and IDS for \ce{CO2}-\ce{CH4}, we performed the \emph{ab initio} calculations ourselves.  Once the \emph{ab initio} data were acquired, the zeroth spectral moment for the system was calculated as 
\begin{equation}
\tilde \Gamma = \frac{32\pi^4}{3 h c}\int_0^\infty \int_\Omega \mu(R,\Omega)^2\E^{-V(R,\Omega)/k_B T}R^2 \D R \D \Omega,
\end{equation}
where $h$ is Planck's constant, $c$ is the speed of light, $R$ is the separation of the molecular centers of mass, $\Omega$ is solid angle, $V$ is the PES, $\mu$ is the IDS, $k_B$ is Boltzmann's constant and $T$ is temperature \citep{Frommhold2006}. 

We assessed the climate effects of the new CIA coefficients using a new iterative line-by-line spectral code \citep{Wordsworth2016b,Schaefer2016}. Using this model allowed us to perform extremely high accuracy globally averaged calculations while spanning a wide range of atmospheric compositions. The code has been validated against a number of analytic results and previous radiative-convective calculations. Further details of our CIA and line-by-line climate calculations are given in the Auxiliary Online Material \citep{Cherepanov2016,Boys1970,Knizia2009,Cohen1965,Clough1992,Rothman2013,Murphy2005,Gruszka1997,Baranov2004,Wordsworth2010,Pierrehumbert2011BOOK,Hansen1974,Claire2012,Beguier2015,Goldblatt2013,Schaefer2016}.

\section{Results}

\begin{figure}[h]
	\begin{center}
		{\includegraphics[width=3.5in]{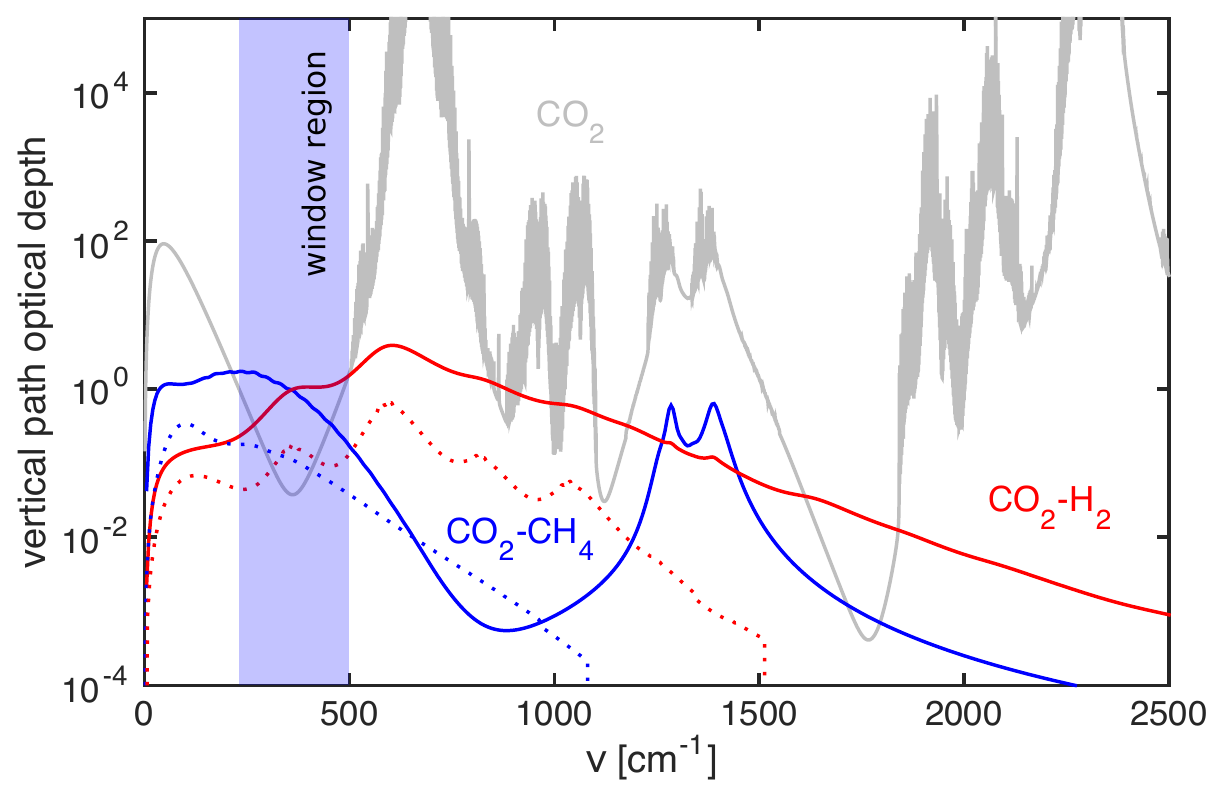}}
	\end{center}
	\caption{Total vertical path optical depth due to \ce{CO2} (gray), \ce{CO2-CH4} CIA (blue) and \ce{CO2-H2} CIA (red) in the early martian atmosphere, assuming a pressure of 1~bar, composition 94\% \ce{CO2}, 3\% \ce{CH4}, 3\% \ce{H2}, and surface temperature of 250~K. Dotted lines show optical depth from CIA when the absorption coefficients of \ce{CO2-H2} and \ce{CO2-CH4} are replaced by those of  \ce{N2-H2} and \ce{N2-CH4}, respectively. Both  the \ce{CO2-H2} and \ce{CO2-CH4} CIA  are strong in a critical window region of the spectrum where absorption by pure \ce{CO2} is weak. }
\label{fig:spectra}
\end{figure}

First, we compared the \ce{CO2}-\ce{H2} and \ce{CO2}-\ce{CH4} CIA coefficients we calculated with previously derived \ce{N2}-\ce{H2} and \ce{N2}-\ce{CH4} CIA data \citep{Borysow1986,Borysow1993,Richard2012}. Figure~\ref{fig:spectra} shows that the peak values of the \ce{CO2} CIA coefficients are significantly stronger than the previously calculated \ce{N2} data. {The difference can be explained by the higher electronegativity of oxygen than carbon, which leads to a more heterogenous electron density distribution for \ce{CO2} than for \ce{N2}. This in turn leads to stronger multipole moments and a higher polarizability, which enhances CIA. For example, the quadrupole moment of \ce{CO2} is approximately 3 times greater than that of \ce{N2} \citep{Graham1998}. A significant portion of CIA scales with the square of the quadrupole moment, leading to a factor of $\sim9$ increase (c.f. the coefficients in Fig.~\ref{fig:spectra}). The \ce{CO2} enhancement } effect is particularly significant for climate because both pairs absorb significantly between 250 and 500~cm$^{-1}$: a key spectral window region for the martian climate \citep{Wordsworth2016b}. 

\begin{figure}[h]
	\begin{center}
		{\includegraphics[width=6in]{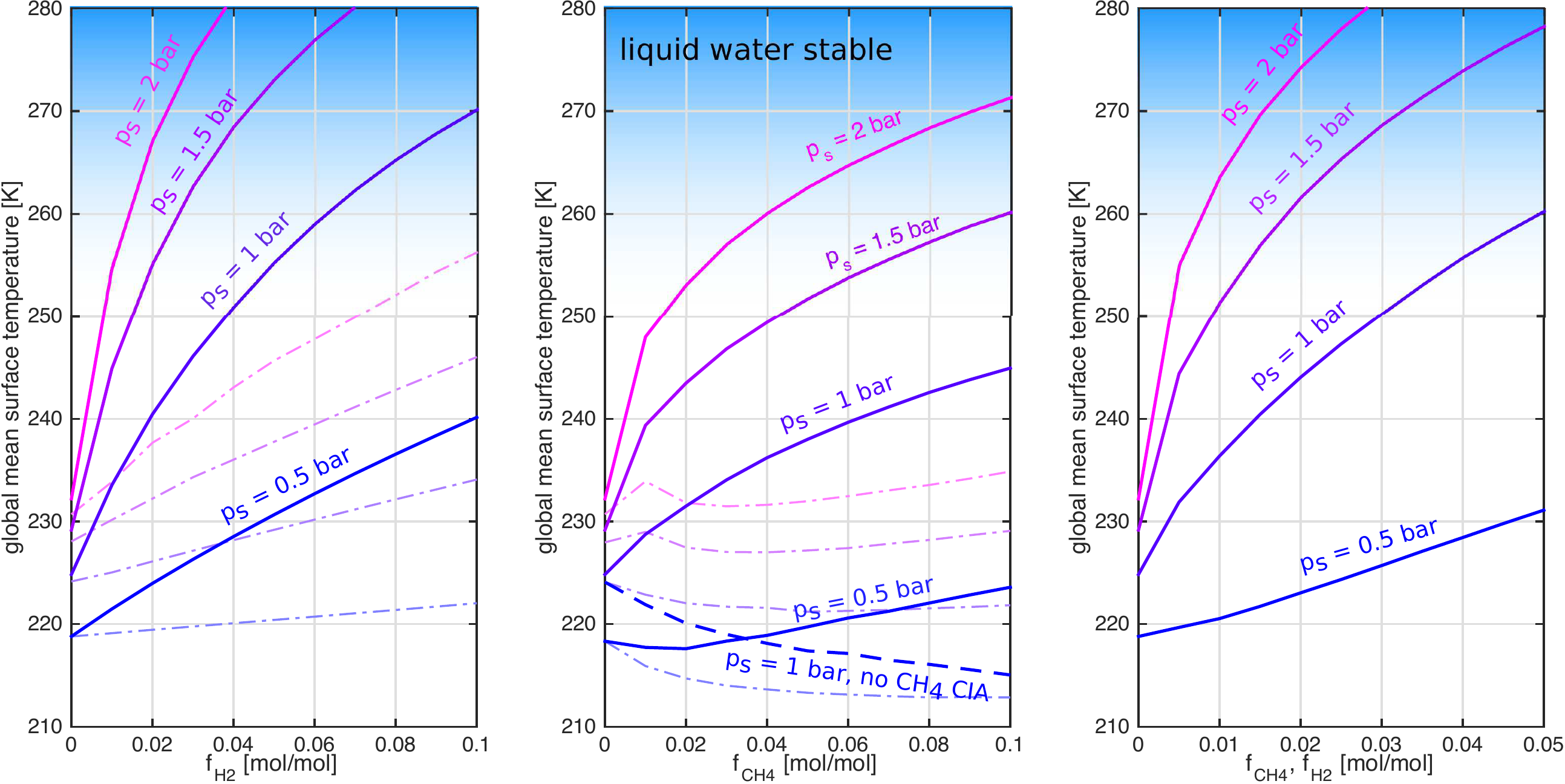}}
	\end{center}
	\caption{Surface temperature in \ce{CO2}-dominated atmospheres as a function of a) \ce{H2} and b) \ce{CH4} molar concentration for various surface pressures $p_s$. The solid lines show results calculated using our new CIA coefficients, while dash-dot lines show results using \ce{N2}-\ce{H2} and \ce{N2}-\ce{CH4} CIA coefficients in place of the correct coefficients. In b), the dashed line shows the case at 1~bar where \ce{CH4} CIA is removed entirely, demonstrating that without it, methane actually has an anti-greenhouse effect. Figure c) shows the case where both \ce{H2} and \ce{CH4} are present in equal amounts. Note the change of scale on the $x$-axis compared to a) and b).}
\label{fig:Tsurf_results}
\end{figure}

These increased opacities translate directly to higher surface temperatures in climate calculations.
Figure~\ref{fig:Tsurf_results}a shows the result of calculating surface temperature using both our new \ce{CO2}-\ce{H2} data and (incorrectly) using \ce{N2}-\ce{H2} as a substitute for \ce{CO2}-\ce{H2}. As can be seen, the difference is significant, with surface temperatures increasing by many tens of degrees for \ce{H2} abundances greater than a few percent. Global mean temperatures exceed 273~K for \ce{H2} molar concentrations from 2.5 to 10\%, depending on the background \ce{CO2} pressure.

Next, we studied the effects of methane. In the past, methane has not been regarded as an effective early martian greenhouse gas because its first vibration-rotation absorption band peaks at 1300~cm$^{-1}$, too far from the blackbody emission spectrum peak at 250-300~K to reduce the outgoing longwave radiation (OLR) significantly \citep{Ramirez2014,Wordsworth2016b}.  Methane also absorbs incoming solar radiation significantly in the near-infrared \citep{Brown2013}.  We find strong \ce{CH4} near-IR absorption, leading to a temperature inversion in the high atmosphere when \ce{CH4} is present. Hence although \ce{CH4} near-IR absorption decreases planetary albedo, its net effect is to slightly \emph{decrease} surface temperatures in the absence of other effects (Fig.~2b).

Despite its anti-greenhouse properties in the near-IR, we nonetheless find that at high abundance, methane can also act as an important greenhouse gas on early Mars. This occurs because the \ce{CO2}-\ce{CH4} CIA absorption peaks in the key 250 to 500~cm$^{-1}$ window region. We find that adding 5\% \ce{CH4} increases global mean temperatures by up to $\sim$30~K, depending on the background \ce{CO2} pressure (Figure~\ref{fig:Tsurf_results}).  Finally, when \ce{CH4} and \ce{H2} are combined in equal proportions, only 3.5\% of each gas is required to achieve 273~K given a 1.5~bar atmosphere (Figure~\ref{fig:Tsurf_results}). Note that 273~K may be an upper limit on the global mean temperature required to explain valley network formation due to the importance of local and seasonal effects in determining runoff (see e.g., \citet{Wordsworth2013a,Kite2013,Rosenberg2015}).

\section{Discussion}

Our spectroscopic CIA and line-by-line climate calculations have shown that a combination of reducing gases in the early martian atmosphere could potentially solve the faint young Sun problem. But is such a solution physically and chemically plausible? While the abundances of methane and hydrogen on Mars today are extremely low \citep{Webster2015}, highly reducing atmospheres are observed elsewhere in the solar system: Titan has a 1.5~bar \ce{N2} dominated atmosphere with \ce{CH4} levels of 4.9\% (mole fraction) near the surface \citep{Niemann2005}. Titan's methane is destroyed by photochemistry on a timescale of order 10~My \citep{Lunine2008}, and is most likely replenished episodically due to destabilization of methane clathrates in the subsurface \citep{Tobie2006}. 

Mars today has a highly oxidized surface and atmosphere due to hydrogen loss to space over geological time. However, early on methane and hydrogen may have been episodically released from the subsurface in quantities sufficient to raise surface temperatures. Serpentinization, a process in which mafic minerals such as olivine are hydrothermally altered to produce reducing gases, has been proposed as the ultimate origin of the \ce{CH4} on Titan \citep{Tobie2006}. Serpentine deposits have been observed on the martian surface at Nili Fossae, Isidis Basin and in some southern highland impact craters \citep{Ehlmann2010}. Extensive serpentinization may also have occurred on early Mars in the deep olivine-rich crust \citep{Chassefiere2013}.  
Study of terrestrial analogs suggests that low-temperature alteration of martian ultramafic rocks would be capable of producing of order $10^{12}-10^{14}$~molecules/cm$^2$/s of \ce{CH4} in local active regions \citep{Etiope2013}. If 5\% of the early martian crust was rich enough in olivine for serpentinization, this translates to a global \ce{CH4} emission rate of $5\times10^{10}-10^{12}$~molecules/cm$^2$/s.


{Volcanism is another source of reduced gases, particularly of \ce{H2}. Hydrogen outgassing is highest if the oxygen fugacity of the early martian mantle was extremely low \citep{Ramirez2014,Batalha2015}. An important problem with volcanism as the sole source of reduced gases, however, is that a mantle reducing enough to outgas sufficient \ce{H2} directly would outgas \ce{CO2} less efficiently, instead retaining large amounts of carbon in the melt \citep{HirschmannWithers2008,Wetzel2013}}. A third potential reduced gas source is \ce{CH4} and \ce{H2} production due to atmospheric thermochemistry following large meteorite impacts. Because peak valley network formation occurs toward the end of the Noachian, a period of higher impact flux than today \citep{Fassett2008,Fassett2011}, this mechanism deserves detailed investigation in future. 

Once outgassed, the primary sinks for \ce{CH4} and \ce{H2} on early Mars would have been chemical destruction of \ce{CH4} and escape of \ce{H2} to space. The lifetime of methane in an atmosphere in which it is abundant is controlled by photodissociation, which is primarily powered by Lyman-$\alpha$ photons (see Auxiliary Online Material). Previous detailed photochemical modeling has shown that this limit is approached in \ce{CO2}-rich atmospheres when $f_{\ce{CH4}} > 0.1-1\%$  \citep{Zahnle1986}. Using an estimate of the solar XUV flux at 3.8~Ga at Mars' semi-major axis as in \citet{Wordsworth2013b} and integrating the solar flux up to 160~nm, the wavelength above which the absorption cross-section of \ce{CH4} becomes negligible \citep{Chen2004}, we calculate an upper limit \ce{CH4} photodestruction rate of $2.5-3.2\times10^{11}$~molecules/cm$^2$/s. This corresponds to a methane residence time of about 250,000~y starting from 5\% \ce{CH4} in a 1.25~bar \ce{CO2} atmosphere. Note that this estimate ignores chemical recycling of dissociated \ce{CH4} in the atmosphere and the decrease in XUV flux due to absorption by escaping hydrogen higher up in the atmosphere, both of which would increase the \ce{CH4} residence time.

The escape of \ce{H2} to space on early Mars would most likely have been limited by diffusion through the homopause, with a characteristic rate of 
\begin{equation}
\Phi_{\ce{H2}} \approx \frac{b_{\ce{CO2-H2}}}{H_{\ce{CO2}}}f_{\ce{H2}}
\end{equation}
where $b_{\ce{CO2-H2}}$ is the \ce{CO2-H2} binary collision coefficient, $H_{\ce{CO2}}$ is the atmospheric scale height, and $f_{\ce{H2}}$ is the hydrogen molar mixing ratio at the homopause. For hydrogen levels of 1-5\% and a homopause temperature range of 150 to 500~K, we find $\Phi_{\ce{H2}}=0.9-6.3\times10^{11}$~molecules/cm$^2$/s: approximately the same magnitude as the maximum rate of \ce{CH4} photolysis. Hence a pulse of \ce{CH4} emission into the early martian atmosphere would result in a mixed \ce{CO2-CH4-H2} composition that would last for a period of 100,000~y or more. {This timescale is more than sufficient to account for the formation of deposits in Gale crater, given the uncertainty range in sedimentation rates \citep{Grotzinger2015}. It is lower than some timescales estimated for valley network formation based on numerical runoff/erosion modeling \citep{Hoke2011}, but is consistent with others \citep{Rosenberg2015}, at least if a high discharge frequency is assumed. Coupled climate and landform evolution modelling in future will be necessary to test whether $\sim10^5$~y formation timescales are indeed sufficient to explain all Noachian fluvial geomorphology.}

What mechanism could cause pulses in reduced gas outgassing rates? One possibility is simply local variations in the geothermal heat flux, which would alter the rate of subsurface aqueous alteration. Another is the contribution of impactors to the atmospheric \ce{H2} and \ce{CH4} inventory. A third possibility is \ce{CH4} clathration \citep{Lasue2015}. Due to adiabatic cooling of the surface under a denser \ce{CO2} atmosphere, most of Mars' surface ice would have stabilized in the southern highlands \citep{Wordsworth2013a}, in the regions where most serpentine has been detected from orbit \citep{Ehlmann2010}. Hence a substantial portion of outgassed methane could have become trapped as clathrate in the cryosphere. Episodic \ce{CH4} release following large perturbations due to volcanism, impacts or obliquity changes would have destabilized clathrates by altering thermal forcing and by sublimation/melting of the overlying ice. Once released, methane and \ce{H2} would cause greenhouse warming, leading to a positive feedback that would destabilize the remaining ice.

Finally, transient \ce{CH4}/\ce{H2} emissions also require \ce{CO2} levels of 0.5~bar or greater to significantly impact surface temperature. From the late Noachian onward, atmospheric \ce{CO2} levels were determined by a balance between volcanic outgassing, escape to space and surface carbonate formation. During this period, coupled modeling of the \ce{^{13}C}/\ce{^{12}C} isotope ratio has constrained Mars' maximum atmospheric pressure to between 1 and 1.8~bar \citep{Hu2015}. While the upper value is a hard limit, the \ce{CO2} pressures we require to cause significant \ce{CH4}/\ce{H2} warming are nonetheless within current evolutionary constraints.

A \ce{CO2}-\ce{CH4} atmosphere on early Mars would not develop a thick haze as on Titan because organic aerosol formation is strongly inhibited for C/O ratios of 0.6 or lower \citep{Zahnle1986,Trainer2006}. However, reaction of atmospheric \ce{CH4} with oxygen from \ce{CO2} photolysis could lead to increased stratospheric \ce{H2O}. This would cause increased formation of high-altitude cirrus clouds, which would enhance warming \citep{Urata2013}, reducing the background \ce{CO2} requirements beyond the baseline calculations shown here. We plan to investigate this possibility in detail in future work.

\section{Conclusion}

We have produced the first physically realistic calculations of reducing greenhouse warming on early Mars. Our results suggest that with just over 1~bar of atmospheric \ce{CO2}, a few percent of \ce{H2} and/or \ce{CH4} would have raised surface temperatures to the point where the hydrological cycle would have been vigorous enough to explain the geological observations. Other effects, particularly the contribution of methane photolysis to cirrus cloud formation, may lower these \ce{CO2} and \ce{H2}/\ce{CH4} abundance requirements further and deserve detailed investigation (probably with a 3D climate model) in future.

Our CIA calculation methodology has been validated against existing data for \ce{N2-H2} and \ce{N2-CH4} pairs. Nonetheless, the complexity of CIA interactions involving \ce{CH4} means that it may not capture all differences between the \ce{N2-CH4} and \ce{CO2-CH4} systems. For this reason we strongly encourage the experimental investigation of \ce{CO2-CH4} CIA in the future. Testing other aspects of the reducing atmosphere scenario for early Mars will require better constraints on the rate of crustal \ce{H2}/\ce{CH4} production during the Noachian and the nature of the early water cycle.  Future investigation of the detailed chemical composition of the martian crust and mantle, along with a continued search for  serpentine and other hydrated minerals, will be important to make further progress.

Besides early Mars, our results have implications for exoplanet habitability and the search for biosignatures. Current definitions of the outer edge of the habitable zone rely on either \ce{CO2} or \ce{H2} and assume that a biosphere would have a detrimental effect on habitability via methanogenic consumption of these gases \citep[e.g.,][]{Pierrehumbert2011}. However, the apparent strength of \ce{CO2}-\ce{CH4} CIA means that an inhabited planet could potentially retain a stable climate at great distances from its host star.

\begin{acknowledgments}
R.~W. acknowledges financial support from the Kavli Foundation and discussions with T.~Robinson and F.~Ding on line-by-line radiative calculations and K.~Zahnle on atmospheric chemistry. S.~L., Y.~K., and A.~V. gratefully acknowledge partial support of this work from RFBR Grants 15-03-03302 and 15-05-00736 and the Russian Academy of Sciences Program 9. B.L.E. thanks B. Sherwood-Lollar and G. Etiope for discussion of \ce{H2}/\ce{CH4} observed in terrestrial serpentinizing systems. The \emph{ab initio} calculations were performed using the HPC resources of the FAS Research Computing Cluster (Harvard University) and the ``Lomonosov'' (Moscow State University) supercomputer. The CIA data produced from our spectroscopic calculations and the line-by-line and temperature data produced from our climate model are available from the lead author on request (\emph{rwordsworth@seas.harvard.edu}).
\end{acknowledgments}


\begin{thebibliography}{61}
\providecommand{\natexlab}[1]{#1}
\expandafter\ifx\csname urlstyle\endcsname\relax
  \providecommand{\doi}[1]{doi:\discretionary{}{}{}#1}\else
  \providecommand{\doi}{doi:\discretionary{}{}{}\begingroup
  \urlstyle{rm}\Url}\fi

\bibitem[{\textit{{Baranov} et~al.}(2004)\textit{{Baranov}, {Lafferty}, and
  {Fraser}}}]{Baranov2004}
{Baranov}, Y.~I., W.~J. {Lafferty}, and G.~T. {Fraser} (2004), {Infrared
  spectrum of the continuum and dimer absorption in the vicinity of the O2
  vibrational fundamental in O2/CO2 mixtures}, \textit{J. Mol. Spectrosc.},
  \textit{228}, 432--440, \doi{10.1016/j.jms.2004.04.010}.

\bibitem[{\textit{Batalha et~al.}(2015)\textit{Batalha, Domagal-Goldman,
  Ramirez, and Kasting}}]{Batalha2015}
Batalha, N., S.~D. Domagal-Goldman, R.~Ramirez, and J.~F. Kasting (2015),
  Testing the early {M}ars {H2--CO2} greenhouse hypothesis with a 1-{D}
  photochemical model, \textit{Icarus}, \textit{258}, 337--349.

\bibitem[{\textit{B{\'e}guier et~al.}(2015)\textit{B{\'e}guier, Liu, and
  Campargue}}]{Beguier2015}
B{\'e}guier, S., A.~W. Liu, and A.~Campargue (2015), An empirical line list for
  methane near 1~$\mu$m (9028 --10,435 cm-1), \textit{Journal of Quantitative
  Spectroscopy and Radiative Transfer}, \textit{166}, 6--12.

\bibitem[{\textit{Borysow and Frommhold}(1986)}]{Borysow1986}
Borysow, A., and L.~Frommhold (1986), Theoretical collision-induced
  rototranslational absorption spectra for modeling {T}itan's atmosphere:
  {H2}-{N2} pairs, \textit{The Astrophysical Journal}, \textit{303}, 495--510.

\bibitem[{\textit{Borysow and Tang}(1993)}]{Borysow1993}
Borysow, A., and C.~Tang (1993), Far infrared {CIA} spectra of {N2-CH4} pairs
  for modeling of {T}itan's atmosphere, \textit{Icarus}, \textit{105}(1),
  175--183.

\bibitem[{\textit{Boys and Bernardi}(1970)}]{Boys1970}
Boys, S.~F., and F.~Bernardi (1970), The calculation of small molecular
  interactions by the differences of separate total energies. some procedures
  with reduced errors, \textit{Molecular Physics}, \textit{19}(4), 553--566.

\bibitem[{\textit{Brown et~al.}(2013)\textit{Brown, Sung, Benner, Devi, Boudon,
  Gabard, Wenger, Campargue, Leshchishina, Kassi et~al.}}]{Brown2013}
Brown, L.~R., K.~Sung, D.~C. Benner, V.~M. Devi, V.~Boudon, T.~Gabard,
  C.~Wenger, A.~Campargue, O.~Leshchishina, S.~Kassi, et~al. (2013), {M}ethane
  line parameters in the {HITRAN2012} database, \textit{Journal of Quantitative
  Spectroscopy and Radiative Transfer}, \textit{130}, 201--219.

\bibitem[{\textit{Chassefi{\`e}re et~al.}(2013)\textit{Chassefi{\`e}re,
  Langlais, Quesnel, and Leblanc}}]{Chassefiere2013}
Chassefi{\`e}re, E., B.~Langlais, Y.~Quesnel, and F.~Leblanc (2013), The fate
  of early {M}ars' lost water: the role of serpentinization, \textit{Journal of
  Geophysical Research: Planets}, \textit{118}(5), 1123--1134.

\bibitem[{\textit{Chen and Wu}(2004)}]{Chen2004}
Chen, F.~Z., and C.~Y.~R. Wu (2004), {T}emperature-dependent photoabsorption
  cross sections in the {VUV-UV} region. i. {M}ethane and ethane. {M}ethane and
  ethane, \textit{Journal of Quantitative Spectroscopy and Radiative Transfer},
  \textit{85}(2), 195--209.

\bibitem[{\textit{Cherepanov et~al.}(2016)\textit{Cherepanov, Kalugina, and
  Buldakov}}]{Cherepanov2016}
Cherepanov, V.~N., Y.~N. Kalugina, and M.~A. Buldakov (2016),
  \textit{Interaction-induced electric properties of van der {W}aals omplexes},
  Springer.

\bibitem[{\textit{Claire et~al.}(2012)\textit{Claire, Sheets, Cohen, Ribas,
  Meadows, and Catling}}]{Claire2012}
Claire, M.~W., J.~Sheets, M.~Cohen, I.~Ribas, V.~S. Meadows, and D.~C. Catling
  (2012), The evolution of solar flux from 0.1 nm to 160 $\mu$m: quantitative
  estimates for planetary studies, \textit{The Astrophysical Journal},
  \textit{757}(1), 95.

\bibitem[{\textit{Clough et~al.}(1992)\textit{Clough, Iacono, and
  Moncet}}]{Clough1992}
Clough, S.~A., M.~J. Iacono, and J.-L. Moncet (1992), Line-by-line calculations
  of atmospheric fluxes and cooling rates: Application to water vapor (paper
  92jd01419), \textit{Journal of Geophysical Research}, \textit{97}, 15--761.

\bibitem[{\textit{Cohen and Roothaan}(1965)}]{Cohen1965}
Cohen, H.~D., and C.~C.~J. Roothaan (1965), Electric dipole polarizability of
  atoms by the {H}artree---{F}ock method. {I}. {T}heory for closed-shell
  systems, \textit{The Journal of Chemical Physics}, \textit{43}(10), S34--S39.

\bibitem[{\textit{Ehlmann et~al.}(2010)\textit{Ehlmann, Mustard, and
  Murchie}}]{Ehlmann2010}
Ehlmann, B.~L., J.~F. Mustard, and S.~L. Murchie (2010), Geologic setting of
  serpentine deposits on {M}ars, \textit{Geophysical research letters},
  \textit{37}(6).

\bibitem[{\textit{Etiope et~al.}(2013)\textit{Etiope, Ehlmann, and
  Schoell}}]{Etiope2013}
Etiope, G., B.~L. Ehlmann, and M.~Schoell (2013), Low temperature production
  and exhalation of methane from serpentinized rocks on {E}arth: a potential
  analog for methane production on {M}ars, \textit{Icarus}, \textit{224}(2),
  276--285.

\bibitem[{\textit{{Fassett} and {Head}}(2008{\natexlab{a}})}]{Fassett2008a}
{Fassett}, C.~I., and J.~W. {Head} (2008{\natexlab{a}}), {Valley network-fed,
  open-basin lakes on Mars: Distribution and implications for Noachian surface
  and subsurface hydrology}, \textit{Icarus}, \textit{198}, 37--56,
  \doi{10.1016/j.icarus.2008.06.016}.

\bibitem[{\textit{{Fassett} and {Head}}(2008{\natexlab{b}})}]{Fassett2008}
{Fassett}, C.~I., and J.~W. {Head} (2008{\natexlab{b}}), {The timing of martian
  valley network activity: Constraints from buffered crater counting},
  \textit{Icarus}, \textit{195}, 61--89, \doi{10.1016/j.icarus.2007.12.009}.

\bibitem[{\textit{{Fassett} and {Head}}(2011)}]{Fassett2011}
{Fassett}, C.~I., and J.~W. {Head} (2011), {Sequence and timing of conditions
  on early Mars}, \textit{Icarus}, \textit{211}, 1204--1214,
  \doi{10.1016/j.icarus.2010.11.014}.

\bibitem[{\textit{{Forget} and {Pierrehumbert}}(1997)}]{Forget1997}
{Forget}, F., and R.~T. {Pierrehumbert} (1997), {Warming Early Mars with Carbon
  Dioxide Clouds That Scatter Infrared Radiation}, \textit{Science},
  \textit{278}, 1273--1276, \doi{10.1126/science.278.5341.1273}.

\bibitem[{\textit{Forget et~al.}(2013)\textit{Forget, Wordsworth, Millour,
  Madeleine, Kerber, Leconte, Marcq, and Haberle}}]{Forget2013}
Forget, F., R.~D. Wordsworth, E.~Millour, J.-B. Madeleine, L.~Kerber,
  J.~Leconte, E.~Marcq, and R.~M. Haberle (2013), 3D modelling of the early
  martian climate under a denser \ce{CO2} atmosphere: Temperatures and \ce{CO2} ice
  clouds., \textit{Icarus}, \doi{10.1016/j.icarus.2012.10.019}.

\bibitem[{\textit{Frommhold}(2006)}]{Frommhold2006}
Frommhold, L. (2006), \textit{Collision-induced absorption in gases}, Cambridge
  University Press.

\bibitem[{\textit{Goldblatt et~al.}(2013)\textit{Goldblatt, Robinson, Zahnle,
  and Crisp}}]{Goldblatt2013}
Goldblatt, C., T.~D. Robinson, K.~J. Zahnle, and D.~Crisp (2013), Low simulated
  radiation limit for runaway greenhouse climates, \textit{Nature Geoscience},
  \textit{6}(8), 661--667.

\bibitem[{\textit{Graham et~al.}(1998)\textit{Graham, Imrie, and
  Raab}}]{Graham1998}
Graham, C., D.~A. Imrie, and R.~E. Raab (1998), Measurement of the electric
  quadrupole moments of co2, co, n2, cl2 and bf3, \textit{Molecular Physics},
  \textit{93}(1), 49--56.

\bibitem[{\textit{Grotzinger et~al.}(2015)\textit{Grotzinger, Gupta, Malin,
  Rubin, Schieber, Siebach, Sumner, Stack, Vasavada, Arvidson
  et~al.}}]{Grotzinger2015}
Grotzinger, J.~P., S.~Gupta, M.~C. Malin, D.~M. Rubin, J.~Schieber, K.~Siebach,
  D.~Y. Sumner, K.~M. Stack, A.~R. Vasavada, R.~E. Arvidson, et~al. (2015),
  Deposition, exhumation, and paleoclimate of an ancient lake deposit, {G}ale
  crater, {M}ars, \textit{Science}, \textit{350}(6257), aac7575.

\bibitem[{\textit{{Gruszka} and {Borysow}}(1997)}]{Gruszka1997}
{Gruszka}, M., and A.~{Borysow} (1997), {R}oto-translational collision-induced
  absorption of {CO$_2$} for the atmosphere of venus at frequencies from 0 to
  250 cm$^{-1}$, at temperatures from 200 to 800 {K}, \textit{Icarus},
  \textit{129}, 172--177, \doi{10.1006/icar.1997.5773}.

\bibitem[{\textit{Halevy and {Head}}(2014)}]{Halevy2014}
Halevy, I., and J.~W. {Head} (2014), Episodic warming of early mars by
  punctuated volcanism, \textit{Nature Geoscience}, \doi{10.1038/ngeo2293}.
  
\bibitem[{\textit{{Hansen} and {Travis}}(1974)}]{Hansen1974}
{Hansen}, J.~E., and L.~D. {Travis} (1974), {Light scattering in planetary
  atmospheres}, \textit{Space Sci. Rev.}, \textit{16}, 527--610,
  \doi{10.1007/BF00168069}.

\bibitem[{\textit{Hirschmann and Withers}(2008)}]{HirschmannWithers2008}
Hirschmann, M.~M., and A.~C. Withers (2008), Ventilation of {CO2} from a
  reduced mantle and consequences for the early martian greenhouse,
  \textit{Earth and Planetary Science Letters}, \textit{270}(1), 147--155.

\bibitem[{\textit{{Hoke} et~al.}(2011)\textit{{Hoke}, {Hynek}, and
  Tucker}}]{Hoke2011}
{Hoke}, M.~R.~T., B.~M. {Hynek}, and G.~E. Tucker (2011), Formation timescales
  of large martian valley networks, \textit{Earth and Planetary Science
  Letters}, \textit{312}(1), 1--12.

\bibitem[{\textit{Hu et~al.}(2015)\textit{Hu, Kass, Ehlmann, and
  Yung}}]{Hu2015}
Hu, R., D.~M. Kass, B.~L. Ehlmann, and Y.~L. Yung (2015), Tracing the fate of
  carbon and the atmospheric evolution of {M}ars, \textit{Nature
  communications}, \textit{6}.

\bibitem[{\textit{{Hynek} et~al.}(2010)\textit{{Hynek}, {Beach}, and
  {Hoke}}}]{Hynek2010}
{Hynek}, B.~M., M.~{Beach}, and M.~R.~T. {Hoke} (2010), {Updated global map of
  Martian valley networks and implications for climate and hydrologic
  processes}, \textit{Journal of Geophysical Research (Planets)}, \textit{115},
  E09,008, \doi{10.1029/2009JE003548}.

\bibitem[{\textit{{Kasting}}(1991)}]{Kasting1991}
{Kasting}, J.~F. (1991), {CO2 condensation and the climate of early Mars},
  \textit{Icarus}, \textit{94}, 1--13.

\bibitem[{\textit{{Kerber} et~al.}(2015)\textit{{Kerber}, {Forget}, and
  {Wordsworth}}}]{Kerber2015}
{Kerber}, L., F.~{Forget}, and R.~D. {Wordsworth} (2015), Sulfur in the early
  martian atmosphere revisited: Experiments with a 3D global climate model,
  \textit{Icarus}, \doi{10.1016/j.icarus.2015.08.011}.

\bibitem[{\textit{Kite et~al.}(2013)\textit{Kite, Halevy, Kahre, Wolff, and
  Manga}}]{Kite2013}
Kite, E.~S., I.~Halevy, M.~A. Kahre, M.~J. Wolff, and M.~Manga (2013), Seasonal
  melting and the formation of sedimentary rocks on mars, with predictions for
  the gale crater mound, \textit{Icarus}, \textit{223}(1), 181--210.

\bibitem[{\textit{Knizia et~al.}(2009)\textit{Knizia, Adler, and
  Werner}}]{Knizia2009}
Knizia, G., T.~B. Adler, and H.-J. Werner (2009), Simplified {CCSD (T)-F12}
  methods: {T}heory and benchmarks, \textit{The Journal of Chemical Physics},
  \textit{130}(5), 054,104.

\bibitem[{\textit{Lasue et~al.}(2015)\textit{Lasue, Quesnel, Langlais, and
  Chassefi{\`e}re}}]{Lasue2015}
Lasue, J., Y.~Quesnel, B.~Langlais, and E.~Chassefi{\`e}re (2015), Methane
  storage capacity of the early martian cryosphere, \textit{Icarus},
  \textit{260}, 205--214.

\bibitem[{\textit{Li et~al.}(2010)\textit{Li, Roy, and Le~Roy}}]{Li2010}
Li, H., P.-N. Roy, and R.~J. Le~Roy (2010), Analytic {M}orse/long-range
  potential energy surfaces and predicted infrared spectra for {CO2}-{H2},
  \textit{Journal of Chemical Physics}, \textit{132}(21), 214,309.

\bibitem[{\textit{Lunine and Atreya}(2008)}]{Lunine2008}
Lunine, J.~I., and S.~K. Atreya (2008), The methane cycle on {T}itan,
  \textit{Nature Geoscience}, \textit{1}(3), 159--164.

\bibitem[{\textit{Murphy and Koop}(2005)}]{Murphy2005}
Murphy, D.~M., and T.~Koop (2005), Review of the vapour pressures of ice and
  supercooled water for atmospheric applications, \textit{Quarterly Journal of
  the Royal Meteorological Society}, \textit{131}(608), 1539--1565.

\bibitem[{\textit{Niemann et~al.}(2005)\textit{Niemann, Atreya, Bauer,
  Carignan, Demick, Frost, Gautier, Haberman, Harpold, Hunten
  et~al.}}]{Niemann2005}
Niemann, H.~B., S.~K. Atreya, S.~J. Bauer, G.~R. Carignan, J.~E. Demick, R.~L.
  Frost, D.~Gautier, J.~A. Haberman, D.~N. Harpold, D.~M. Hunten, et~al.
  (2005), The abundances of constituents of titan's atmosphere from the {GCMS}
  instrument on the {H}uygens probe, \textit{Nature}, \textit{438}(7069),
  779--784.

\bibitem[{\textit{{Pierrehumbert} and {Gaidos}}(2011)}]{Pierrehumbert2011}
{Pierrehumbert}, R., and E.~{Gaidos} (2011), {Hydrogen Greenhouse Planets
  Beyond the Habitable Zone}, \textit{The Astrophysical Journal Letters},
  \textit{734}, L13, \doi{10.1088/2041-8205/734/1/L13}.

\bibitem[{\textit{Pierrehumbert}(2011)}]{Pierrehumbert2011BOOK}
Pierrehumbert, R.~T. (2011), \textit{Principles of Planetary Climate},
  Cambridge University Press.

\bibitem[{\textit{{Postawko} and {Kuhn}}(1986)}]{Postawko1986}
{Postawko}, S.~E., and W.~R. {Kuhn} (1986), {Effect of the greenhouse gases
  (CO$_{2}$, H$_{2}$O, SO$_{2}$) on martian paleoclimate}, \textit{Journal of
  Geophysical Research}, \textit{91}, 431--D438.

\bibitem[{\textit{Ramirez et~al.}(2014)\textit{Ramirez, Kopparapu, Zugger,
  Robinson, Freedman, and Kasting}}]{Ramirez2014}
Ramirez, R.~M., R.~Kopparapu, M.~E. Zugger, T.~D. Robinson, R.~Freedman, and
  J.~F. Kasting (2014), Warming early {M}ars with {CO$_2$} and {H$_2$},
  \textit{Nature Geoscience}, \textit{7}(1), 59--63.

\bibitem[{\textit{Richard et~al.}(2012)\textit{Richard, Gordon, Rothman, Abel,
  Frommhold, Gustafsson, Hartmann, Hermans, Lafferty, Orton
  et~al.}}]{Richard2012}
Richard, C., I.~Gordon, L.~Rothman, M.~Abel, L.~Frommhold, M.~Gustafsson, J.-M.
  Hartmann, C.~Hermans, W.~Lafferty, G.~Orton, et~al. (2012), New section of
  the {HITRAN} database: {C}ollision-induced absorption ({CIA}),
  \textit{Journal of Quantitative Spectroscopy and Radiative Transfer},
  \textit{113}(11), 1276--1285.

\bibitem[{\textit{Rosenberg and Head}(2015)}]{Rosenberg2015}
Rosenberg, E.~N., and J.~W. Head (2015), Late noachian fluvial erosion on mars:
  Cumulative water volumes required to carve the valley networks and grain size
  of bed-sediment, \textit{Planetary and Space Science}, \textit{117},
  429--435.

\bibitem[{\textit{Rothman et~al.}(2013)\textit{Rothman, Gordon, Babikov, Barbe,
  Benner, Bernath, Birk, Bizzocchi, Boudon, Brown et~al.}}]{Rothman2013}
Rothman, L.~S., I.~E. Gordon, Y.~Babikov, A.~Barbe, D.~C. Benner, P.~F.
  Bernath, M.~Birk, L.~Bizzocchi, V.~Boudon, L.~R. Brown, et~al. (2013), The
  {HITRAN2012} molecular spectroscopic database, \textit{Journal of
  Quantitative Spectroscopy and Radiative Transfer}, \textit{130}, 4--50.

\bibitem[{\textit{{Sagan}}(1977)}]{Sagan1977}
{Sagan}, C. (1977), Reducing greenhouses and the temperature history of {E}arth
  and {M}ars, \textit{Nature}, \doi{10.1038/269224a0}.

\bibitem[{\textit{Schaefer et~al.}(2016)\textit{Schaefer, Wordsworth,
  Berta-Thompson, and Sasselov}}]{Schaefer2016}
Schaefer, L., R.~D. Wordsworth, Z.~Berta-Thompson, and D.~Sasselov (2016),
  Predictions of the atmospheric composition of {GJ}1132b, \textit{The
  Astrophysical Journal}, \textit{829}(2), 63.

\bibitem[{\textit{{Segura} et~al.}(2002)\textit{{Segura}, {Toon}, {Colaprete},
  and {Zahnle}}}]{Segura2002}
{Segura}, T.~L., O.~B. {Toon}, A.~{Colaprete}, and K.~{Zahnle} (2002),
  {Environmental Effects of Large Impacts on Mars}, \textit{Science},
  \textit{298}, 1977--1980.

\bibitem[{\textit{Tobie et~al.}(2006)\textit{Tobie, Lunine, and
  Sotin}}]{Tobie2006}
Tobie, G., J.~I. Lunine, and C.~Sotin (2006), Episodic outgassing as the origin
  of atmospheric methane on {T}itan, \textit{Nature}, \textit{440}(7080),
  61--64.

\bibitem[{\textit{Trainer et~al.}(2006)\textit{Trainer, Pavlov, DeWitt,
  Jimenez, McKay, Toon, and Tolbert}}]{Trainer2006}
Trainer, M.~G., A.~A. Pavlov, H.~L. DeWitt, J.~L. Jimenez, C.~P. McKay, O.~B.
  Toon, and M.~A. Tolbert (2006), Organic haze on {T}itan and the early
  {E}arth, \textit{Proceedings of the National Academy of Sciences},
  \textit{103}(48), 18,035--18,042.

\bibitem[{\textit{Urata and Toon}(2013)}]{Urata2013}
Urata, R.~A., and O.~B. Toon (2013), Simulations of the martian hydrologic
  cycle with a general circulation model: Implications for the ancient martian
  climate, \textit{Icarus}, \doi{10.1016/j.icarus.2013.05.014}.

\bibitem[{\textit{Webster et~al.}(2015)\textit{Webster, Mahaffy, Atreya,
  Flesch, Mischna, Meslin, Farley, Conrad, Christensen, Pavlov
  et~al.}}]{Webster2015}
Webster, C.~R., P.~R. Mahaffy, S.~K. Atreya, G.~J. Flesch, M.~A. Mischna, P.-Y.
  Meslin, K.~A. Farley, P.~G. Conrad, L.~E. Christensen, A.~A. Pavlov, et~al.
  (2015), Mars methane detection and variability at {G}ale crater,
  \textit{Science}, \textit{347}(6220), 415--417.

\bibitem[{\textit{Wetzel et~al.}(2013)\textit{Wetzel, Rutherford, Jacobsen,
  Hauri, and Saal}}]{Wetzel2013}
Wetzel, D.~T., M.~J. Rutherford, S.~D. Jacobsen, E.~H. Hauri, and A.~E. Saal
  (2013), Degassing of reduced carbon from planetary basalts,
  \textit{Proceedings of the National Academy of Sciences}, \textit{110}(20),
  8010--8013.

\bibitem[{\textit{Wordsworth and Pierrehumbert}(2013)}]{Wordsworth2013c}
Wordsworth, R., and R.~Pierrehumbert (2013), Hydrogen-nitrogen greenhouse
  warming in {E}arth's early atmosphere, \textit{Science}, \textit{339}(6115),
  64--67, \doi{10.1126/science.1225759}.

\bibitem[{\textit{{Wordsworth} et~al.}(2010)\textit{{Wordsworth}, {Forget}, and
  {Eymet}}}]{Wordsworth2010}
{Wordsworth}, R., F.~{Forget}, and V.~{Eymet} (2010), {Infrared
  collision-induced and far-line absorption in dense CO2 atmospheres},
  \textit{Icarus}, \textit{210}, 992--997, \doi{10.1016/j.icarus.2010.06.010}.

\bibitem[{\textit{Wordsworth et~al.}(2013)\textit{Wordsworth, Forget, Millour,
  Head, Madeleine, and Charnay}}]{Wordsworth2013a}
Wordsworth, R., F.~Forget, E.~Millour, J.~W. Head, J.-B. Madeleine, and
  B.~Charnay (2013), Global modelling of the early martian climate under a
  denser {CO2} atmosphere: Water cycle and ice evolution, \textit{Icarus},
  \textit{222}(1), 1--19.

\bibitem[{\textit{Wordsworth}(2016)}]{Wordsworth2016b}
Wordsworth, R.~D. (2016), The climate of early {M}ars, \textit{Annual Review of
  Earth and Planetary Sciences}, \textit{44}(1).

\bibitem[{\textit{{Wordsworth} and {Pierrehumbert}}(2013)}]{Wordsworth2013b}
{Wordsworth}, R.~D., and R.~T. {Pierrehumbert} (2013), Water loss from
  terrestrial planets with {CO2}-rich atmospheres, \textit{The Astrophysical
  Journal}, \textit{778}(2), 154.

\bibitem[{\textit{Zahnle}(1986)}]{Zahnle1986}
Zahnle, K.~J. (1986), Photochemistry of methane and the formation of
  hydrocyanic acid ({HCN}) in the {E}arth's early atmosphere, \textit{Journal
  of Geophysical Research: Atmospheres (1984--2012)}, \textit{91}(D2),
  2819--2834.

\end{thebibliography}

\end{article}

\newpage

\begin{figure}[h]
	\begin{center}
		{\includegraphics[width=3.5in]{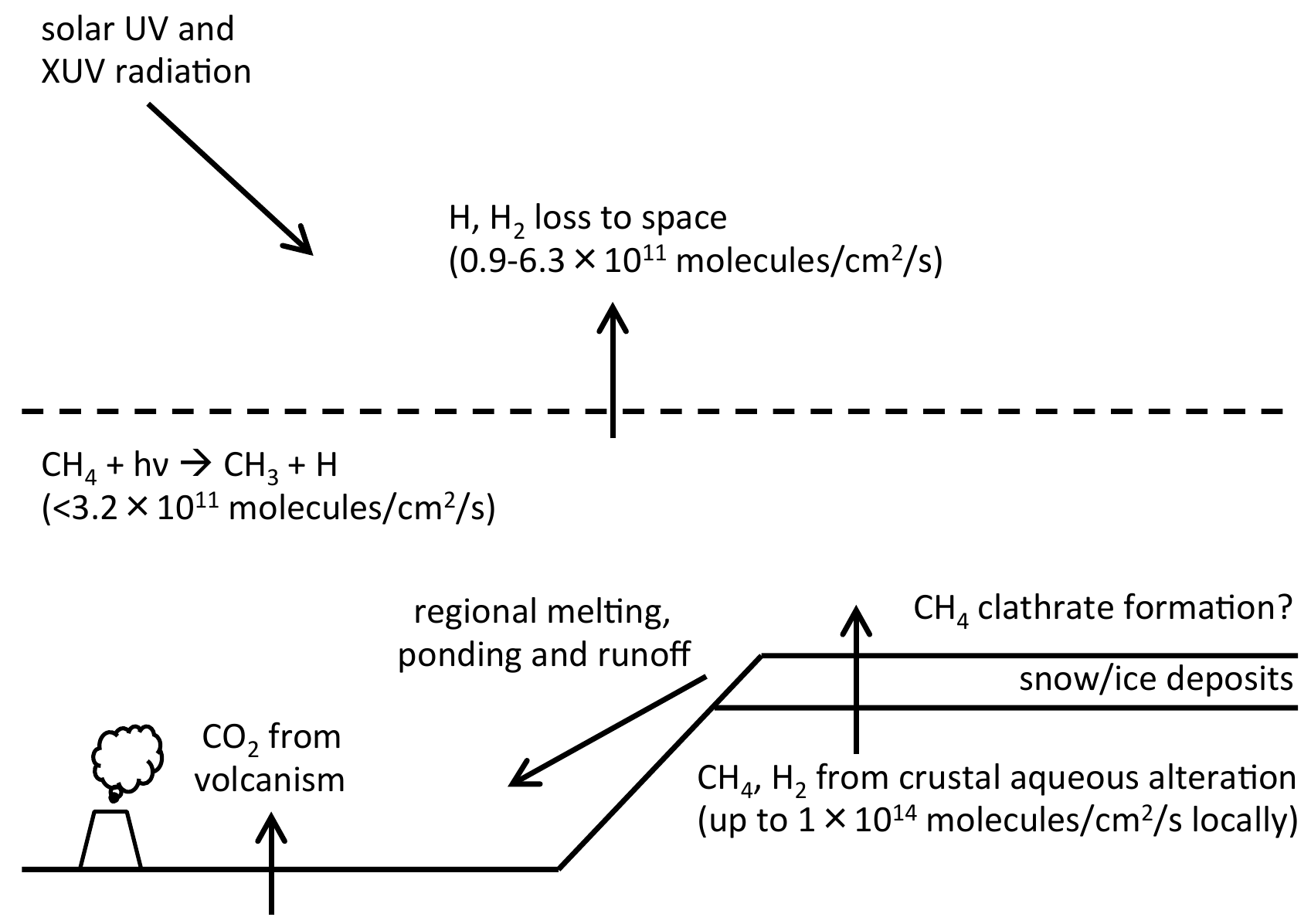}}
	\end{center}
	\caption{Schematic of key processes on early Mars in the transient reducing atmosphere scenario. Highland ice deposits created by adiabatic cooling under a denser \ce{CO2} atmosphere are episodically melted by \ce{H2}/\ce{CH4} warming, leading to runoff, lake formation and fluvial erosion.}
\label{fig:schematic}
\end{figure}

\end{document}